\documentclass[runningheads]{llncs}
\usepackage[utf8]{inputenc}
\usepackage[english]{babel}
\usepackage{amsmath}
\usepackage{amsfonts}
\usepackage{amssymb}
\usepackage{graphicx}
\usepackage{mathtools}
\usepackage[mathletters]{ucs}
\usepackage{multirow}
\DeclarePairedDelimiter{\ceil}{\lceil}{\rceil}

\begin{document}

\title{Deep Learning for ECG Segmentation}

\author{Viktor Moskalenko\orcidID{0000-0003-0309-7680} \and
Nikolai~Zolotykh\orcidID{0000-0003-4542-9233} \and
Grigory~Osipov\orcidID{0000-0003-2841-8399}}

\authorrunning{V. Moskalenko et al.}

\institute{Lobachevsky University of Nizhni Novgorod, \\ Gagarin ave. 23, Nizhni Novgorod 603950 Russia
\email{\{viktor.moskalenko,nikolai.zolotykh,grigory.osipov\}@itmm.unn.ru}}

\maketitle              

\begin{abstract}
We propose an algorithm for electrocardiogram (ECG) segmentation using a UNet-like full-convolutional neural network. 
The algorithm receives an arbitrary sampling rate ECG signal as an input, 
and gives a list of onsets and offsets of P and T waves and QRS complexes as output.
Our method of segmentation differs from others in speed, a small number of parameters and a good generalization: it is adaptive to different sampling rates and it is generalized to various types of ECG monitors.
The proposed approach is superior to other state-of-the-art segmentation methods in terms of quality. 
In particular, $F$1-measures for detection of onsets and offsets
of P and T waves and for QRS-complexes are at least $97.8$\%, $99.5$\%, and $99.9$\%, respectively.

\end{abstract}

\section{Introduction}

The electrocardiogram (ECG) is a recording of the electrical activity of the heart, obtained with the help of electrodes located on the human body. This is one of the most important methods for the diagnosis of heart diseases. The ECG is usually treated by a doctor. Recently, automatic ECG analysis is of great interest. 

\begin{figure}
  \includegraphics[width=\linewidth]{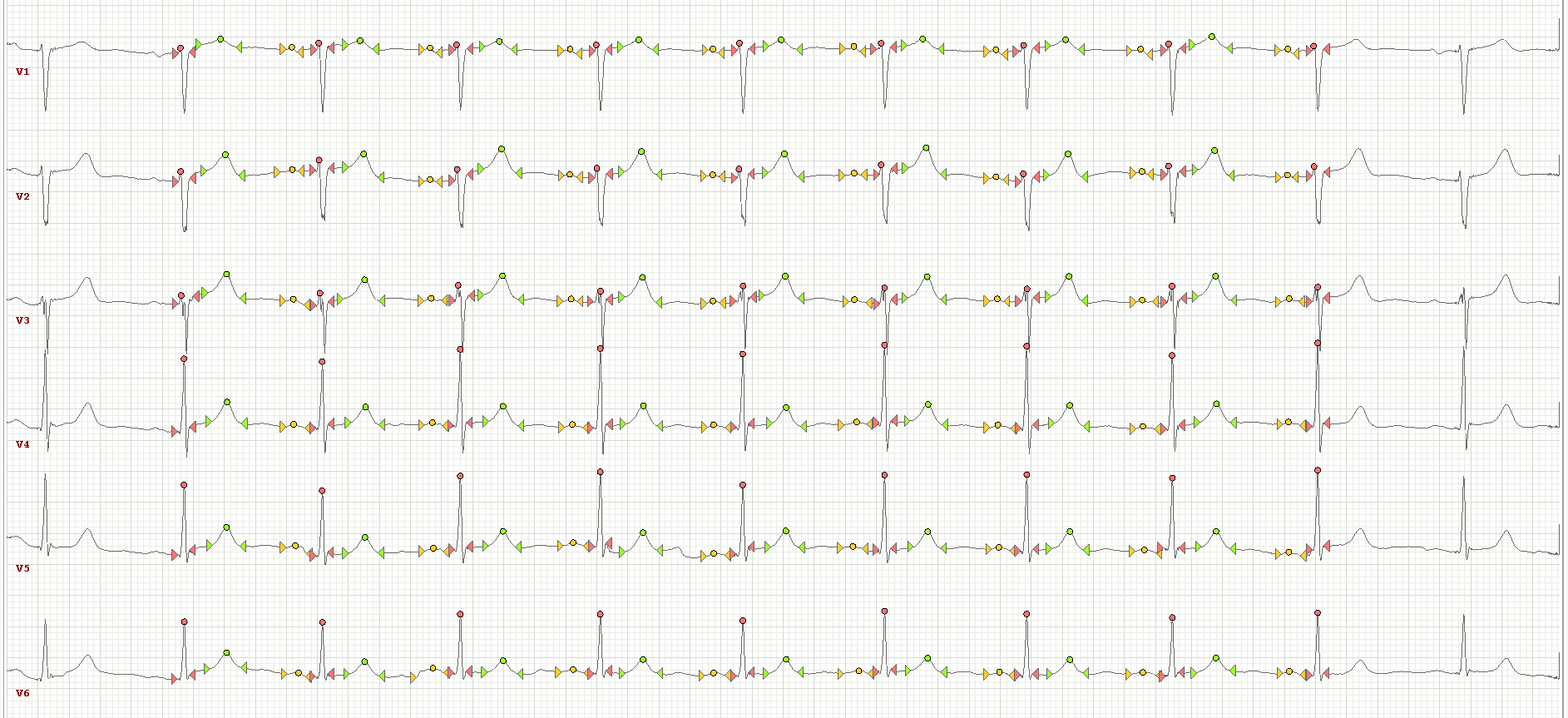}
  \caption{An example of medical segmentation. Yellow color corresponds to P waves, red to QRS complexes, green to T waves. 
	The symbol $\rhd$ means the onset of a wave, $\circ$ means the wave peak, $\lhd$ corresponds to the offset of a wave.}
  \label{fig:delineationDoc}
\end{figure}

The ECG analysis includes detection of QRS complexes, P and T waves, followed by an analysis of their shapes, amplitudes, relative positions, etc. (see Fig.\,\ref{fig:delineationDoc}). The detection of onsets and offsets of QRS complexes and P and T waves is also called segmentation or delineation of the ECG signal.

Accurate ECG automatic segmentation is a difficult problem for the following reasons. 
For example, the P wave has a small amplitude and can be difficult to identify due to interference arising from the movement of electrodes, muscle noise, etc. P and T waves can be biphasic, which makes it difficult to accurately determine their onsets and offsets. Some cardiac cycles may not contain all standard segments, for example, the P wave may be missing, etc. 

Among the methods of automatic ECG segmentation, methods using wavelet transforms have proven to be the best
\cite{Bote2017,Kalyakulina2018,Li1995,DiMarco2011,Martinez2010,Rincon2011}. 
In \cite{Sereda2018}, a neural network approach for ECG segmentation is proposed. 
The segmentation quality turned out to be close to the quality obtained by state-of-the-art algorithms based on wavelet transform, but still, as a rule, lower.
In this paper, we suggest using the UNet-like \cite{UNet2015} neural network.
As a result, using the neural network approach, it is possible to achieve and even exceed the quality of segmentation obtained by other algorithms.
In terms of quality, the proposed approach is superior to analogues. 
In particular, $F$1-measures for detection of onsets and offsets
of P and T waves and for QRS-complex are at least $97.8$\%, $99.5$\%, and $99.9$\%, respectively.

In addition, the proposed segmentation method differs from analogous in speed, a small number of parameters in the neural network and good generalization: it is adaptive to different sampling rates and is generalized to various types of ECG monitors.

The main differences of the proposed approach from the paper \cite{Sereda2018} follow:
\begin{itemize}
	\item in \cite{Sereda2018}, an ensemble of $12$ convolutional neural networks is used; 
	      here we use one full-convolutional neural network with skip links;
  \item in contrast to the present work, \cite{Sereda2018} does not use postprocessing;
	\item in \cite{Sereda2018}, a preprocessing is used to remove a isoline drift; we process signals as is; 
	      in Section \ref{sec:Examples}, we will see that the quality of ECG segmentation is high even in the case of the isoline drift.				
\end{itemize}

\section{Algorithm}

\subsection{Preprocessing}
%

The neural network described below was trained on a dataset of ECG signals with the sampling frequency 500 Hz and the duration 10 s
(see Section \ref{section:LUDB}).
In order to use this network for signals of a different frequency or/and a different duration, we propose the following preprocessing.
Let the frequency of an input signal $x = (x_1, x_2, \dots, x_n)$ be $\nu$, and the network is trained on signals with the frequency $\mu$. Then $T=n/\nu$ is the signal duration.
Convert the input signal as follows.

\begin{enumerate}
	\item[1.] Form an array of time samples $t = (t_1, t_2, \dots, t_n)$,
	          where $t_i = \dfrac{(2i-1)T}{2n}$ are the midpoints of the time intervals formed by dividing the segment $[0, T]$ into $n$ 
						equal parts	$(i=1,2,\dots,n)$. 

	\item[2.] On the set of points $\{(t_1, x_1), (t_2, x_2), \dots, (t_n, x_n) \}$, construct the cubic spline \cite{Spline}.

	\item[3.] Form the array of new time samples $t'=(t_1', t_2', \dots, t_m')$, where
            $$
              m=\ceil[\big]{\mu T}, \qquad t_i'=\frac{(2i-1)T}{2m}.
            $$

	\item[4.] Using the cubic spline, find the signal values at $t'$. The resulting array will be the input to the neural network.
\end{enumerate}

\subsection{The neural network architecture}

\begin{figure}
  \includegraphics[width=\linewidth]{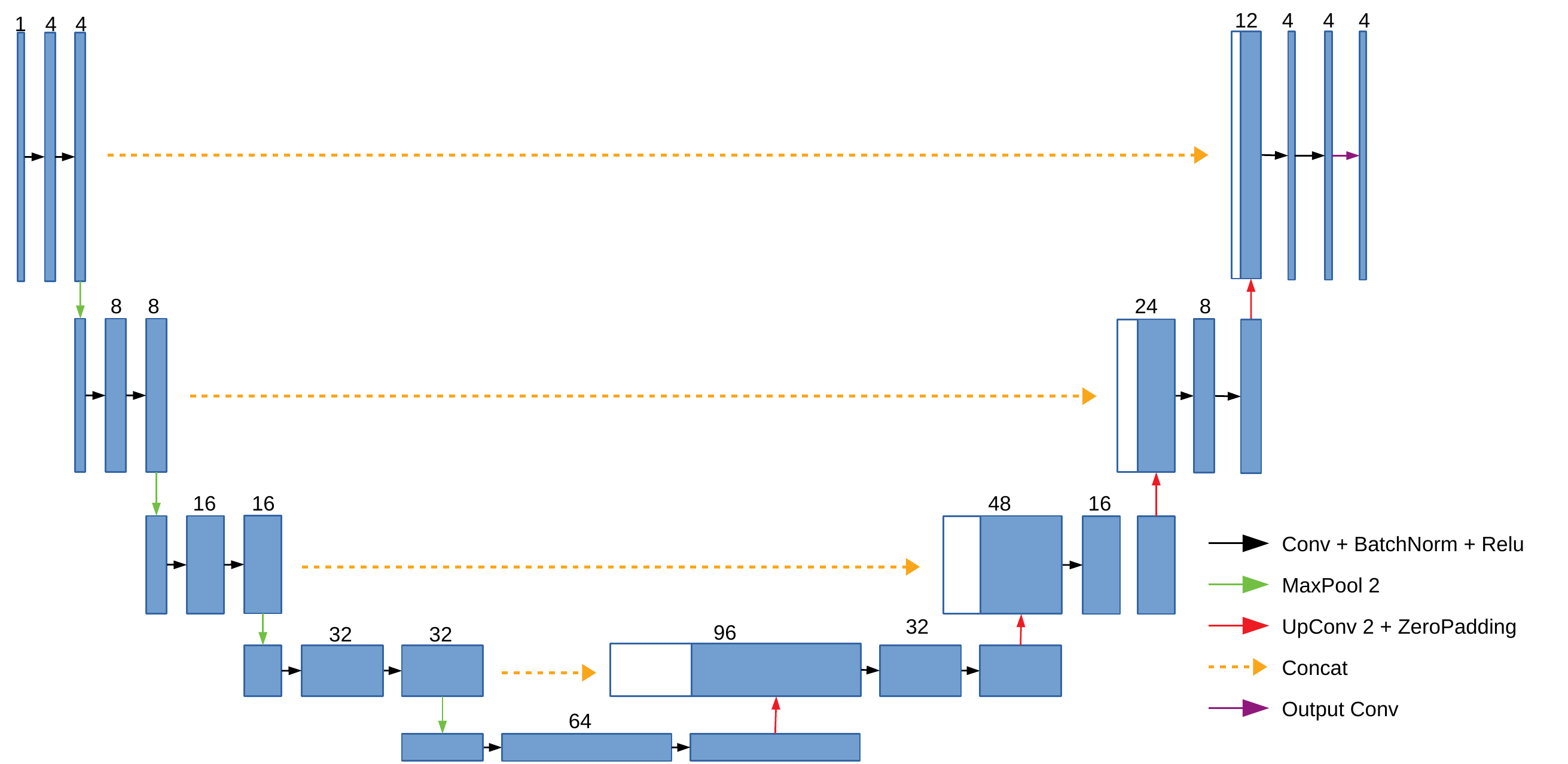}
  \caption{Neural network architecture}
  \label{fig:modelPic}
\end{figure}

The architecture of the neural network (see Fig.\,\ref{fig:modelPic}) is similar to the UNet architecture \cite{UNet2015}. 
The input of the neural network is a vector of length $l$, where $l$ is the length of the ECG signal received from one lead.
Each lead is fed to the input of the neural network separately.

The output size is $(4, l)$. 
Each column of the output matrix contains $4$ scores, that characterize the confidence degree of the neural network that the current value of the signal belongs to the segments P, QRS, T or none of the above. 
The proposed neural network includes the following layers:

\begin{itemize}
	\item[(i)]   $4$ blocks, each of which includes two convolutional layers with batch normalization and the Relu activation function; 
	             these blocks are connected sequentially with MaxPooling layers;
	\item[(ii)]  the output from the previous layer through the MaxPooling layer is fed to the input of another block containing two
	             convolutional layers with batch normalization and the Relu activation function;
	\item[(iii)] the output from the previous layer through the deconvolution and zero padding layers is concatenated with the output 
	             from the layer (ii) and is fed to the input of the block that includes two convolutional layers each with 
							 the batch normalization and the Relu activation function; 
	\item[(iv)]  the output from the previous layer through the deconvolution and zero padding layers is sequentially fed to the input of
	             another $4$ blocks containing two convolutional layers each with batch normalization and Relu activation function; 
							 each time the output is concatenated with the output from the corresponding layers (i) in the reverse order;
	\item[(v)]   the output from the previous layer is fed to the input of another convolutional layer.
\end{itemize}

%
%

All convolutional layers have the following characteristics: $\text{\it kernel-size}=9$, $\text{\it padding}=4$.
All deconvolution layes have $\text{\it kernel-size}=8, \text{\it stride}=2, \text{\it padding}=3$.
For the last convolutional layer $\text{\it kernel-size}=1$.

The main differences between the proposed network and UNet follow:
\begin{itemize}
	\item we use 1d convolutions instead of 2d convolutions;
	\item we use a different number of channels and different parameters in the convolutions;
	\item we use of copy $+$ zero pad layers instead of copy $+$ crop layers;
	      as a result, in the proposed method the dimension of the output is the same as the input;
				in contrast, at the output of the UNet network, we obtain a segmentation of only a part of the image.
\end{itemize}

\subsection{Postprocessing}

The output of the neural network is the matrix of size $(4, l)$, where $l$ is the input signal length. 
Applying the argmax function to the columns of the matrix, 
we obtain a vector of length $l$. 
Form an array of waves, finding all continuous segments with the same label.



For processing multi-leads ECG (a typical number of leads is $12$), we propose to process each lead independently, 
and then find the average of the resulting scores.
As we will see in the Section \label{sec: Experiments}, such an analysis improves the quality of the prediction.

\section{Experimental results}

\subsection{LUDB dataset} \label{section:LUDB}

The training of the neural network and experiment conducting were performed on the extended LUDB dataset \cite{LUDB2018}.
The dataset consists of a $455$ $12$-leads ECG with the duration of $10$ seconds recorded with the sampling rate of $500$ Hz.
For comparison of algorithms, the dataset was divided into a train and a test sets, 
where the test consists of $200$ ECG signals borrowed from the original LUDB dataset.
Since the proposed neural network elaborate the leads independently, 
$255 \times 12 = 3060$ signals of length $500 \times 10 = 5000$ were used for training.
To prevent overfitting, augmentation of data was performed: 
at each batch iteration, a random continuous ECG fragment of $4$ seconds was fed to the input of the neural network.

The LUDB dataset has the following feature.
One (sometimes two) first and last cardiac cycles are not annotated. 
At the same time, the first and last marked segments are necessarily QRS 
(see an exanmple in Fig. \ref{fig:delineationDoc}). 
To implement a correct comparison with the reference segmentation, the following modifications were made in the algorithm:
\begin{itemize}
	\item during augmentation, the first and last $2$ seconds were not taken, 
	      i.\,e. subsequences of the length of $4$ seconds were chosen starting from the $2$-nd to the $4$-th 
				(ending from the $6$-th to the $8$-th seconds);
	\item in order to avoid a large number of false positives, the first and the last cardiac cycles were removed 
	      during the validation of the algorithm.
\end{itemize}

\subsection{Comparison of the algorithms} \label{sec:Experiments}

Table \ref{table:Compare} contains results of the experiment and the comparison of the results with 
one of the best segmentation algorithm using wavelets \cite{Kalyakulina2018}
and the neural network segmentation algorithm \cite{Sereda2018}.
The last line shows the characteristics of our algorithm that analyses the leads independently 
for a test set consisting of $200 \times 12 = 2400$ ECG.

The quality of the algorithms is determined using the following procedure.
According to the recommendations of the Association for Medical Instrumentation \cite{AAMI1999}, it is considered that an onset or an offset are detected correctly, if their deviation from the doctor annotations does not exceed in absolute value the {\it tolerance} of 150 ms.

If an algorithm correctly detects a significant point (an onset or an offset of one of the P, QRS, T segments), 
then a true positive result (TP) is counted and the time deviation (error) of the automatic determined point from the manually marked point is measured.
If there is no corresponding significant point in the test sample in the neighborhood of $\pm \text{\it tolerance}$ of the detected significant point, then the I type error is counted (false positive -- FP).
If the algorithm does not detect a significant point, then the II type error is counted (false negative -- FN).

Following \cite{Bote2017,DiMarco2011,Martinez2010,Rincon2011},
we measure the following quality metrics:
\begin{itemize}
	\item the mean error $m$;
  \item the standard deviation $\sigma$ of the mean error;
  \item the sensitivity, or recall, $\text{\it Se} = \text{\rm TP}/(\text{\rm TP} + \text{\rm FN})$;
  \item the positive predictive value, or precision, $\text{\it PPV} = \text{\rm TP}/(\text{\rm TP} + \text{\rm FP})$.
\end{itemize}	
Here TP, FP, FN denotes the total number of correct solutions, type I errors, and
type II errors, respectively.
We also give the value of
\begin{itemize}
  \item the $F1$-measure: $F1 = 2\,\dfrac{\text{\it Se}\cdot\text{\it PPV}}{\text{\it Se}+\text{\it PPV}}$.
\end{itemize}

\begin{table}
	\begin{center}
		\resizebox{\textwidth}{!}{%
			\begin{tabular}{|p{2.8cm}|c|c|c|c|c|c|c|}
				\hline
				\multicolumn{2}{|c|}{     } & P onset & P offset & QRS onset & QRS offset & T onset & T offset \\

				\hline
				\begin{tabular}{l} Kalyakulina \\ \textit{et al.} \cite{Kalyakulina2018} \end{tabular}
				& \begin{tabular}{c}{\it Se} (\%) \\ {\it PPV} (\%) \\ {\it F}1 (\%) \\ $m\pm\sigma(ms)$ \end{tabular} 
				& \begin{tabular}{c}$98.46$ \\ $96.41$ \\ $97.42$ \\ $-2.7 \pm10.2$ \end{tabular} 
				& \begin{tabular}{c}$98.46$ \\ $96.41$ \\ $97.42$ \\ $0.4\pm11.4$ \end{tabular} 
				& \begin{tabular}{c}$99.61$ \\ $99.87$ \\ $99.74$ \\ $-8.1\pm7.7$ \end{tabular} 
				& \begin{tabular}{c}$99.61$ \\ $99.87$ \\ $99.74$ \\ $3.8\pm8.8$ \end{tabular} 
				& \begin{tabular}{c}$-$ \end{tabular} 
				& \begin{tabular}{c}$98.03$ \\ $98.84$ \\ $98.43$ \\ $5.7\pm 15.5$ \end{tabular} \\

				\hline
				\begin{tabular}{l} ~ \\ Sereda \textit{et al.} \cite{Sereda2018} \\ ~ \end{tabular}
				& \begin{tabular}{c}{\it Se} (\%) \\ {\it PPV} (\%) \\ {\it F}1 (\%) \\ $m\pm\sigma(ms)$ \end{tabular} 
				& \begin{tabular}{c}$95.20$ \\ $82.66$ \\ $88.49$ \\ $2.7\pm21.9$ \end{tabular} 
				& \begin{tabular}{c}$95.39$ \\ $82.59$ \\ $88.53$ \\ $-7.4\pm28.6$ \end{tabular} 
				& \begin{tabular}{c}$99.51$ \\ $98.17$ \\ $98.84$ \\ $2.6\pm12.4$ \end{tabular} 
				& \begin{tabular}{c}$99.50$ \\ $97.96$ \\ $98.72$ \\ $-1.7\pm14.1$ \end{tabular} 
				& \begin{tabular}{c}$97.95$ \\ $94.81$ \\ $96.35$ \\ $8.4\pm28.2$ \end{tabular} 
				& \begin{tabular}{c}$97.56$ \\ $94.96$ \\ $96.24$ \\ $-3.1\pm28.2$ \end{tabular} \\
				
				\hline
				\begin{tabular}{l} ~ \\ This work \\ ~ \end{tabular}
				& \begin{tabular}{c}{\it Se} (\%) \\ {\it PPV} (\%) \\ {\it F}1 (\%) \\ $m\pm\sigma(ms)$ \end{tabular} 
				& \begin{tabular}{c}$98.05$ \\ $97.73$ \\ $\bf 97.89$ \\ $-0.6\pm17.5$ \end{tabular} 
				& \begin{tabular}{c}$98.01$ \\ $97.69$ \\ $\bf 97.85$ \\ $-2.4\pm18.4$ \end{tabular} 
				& \begin{tabular}{c}$100.00$ \\ $99.93$ \\ $99.97$ \\ $1.5\pm11.1$ \end{tabular} 
				& \begin{tabular}{c}$100.00$ \\ $99.93$ \\ $99.97$ \\ $2.0\pm10.6$ \end{tabular} 
				& \begin{tabular}{c}$99.68$ \\ $99.37$ \\ $\bf 99.52$ \\ $2.9\pm23.7$ \end{tabular} 
				& \begin{tabular}{c}$99.77$ \\ $99.46$ \\ $\bf 99.61$ \\ $-2.4\pm30.4$ \end{tabular} \\
				
				\hline
				\begin{tabular}{l} This work \\ (only lead II)  \end{tabular}
				& \begin{tabular}{c}{\it Se} (\%) \\ {\it PPV} (\%) \\ {\it F}1 (\%) \\ $m\pm\sigma(ms)$ \end{tabular} 
				& \begin{tabular}{c}$98.61$ \\ $95.61$ \\ $97.09$ \\ $-4.1\pm20.4$ \end{tabular} 
				& \begin{tabular}{c}$98.59$ \\ $95.59$ \\ $97.07$ \\ $3.7\pm19.6$ \end{tabular} 
				& \begin{tabular}{c}$99.99$ \\ $99.99$ \\ $\bf 99.99$ \\ $1.8\pm13.0$ \end{tabular} 
				& \begin{tabular}{c}$99.99$ \\ $99.99$ \\ $\bf 99.99$ \\ $-0.2\pm11.4$ \end{tabular} 
				& \begin{tabular}{c}$99.32$ \\ $99.02$ \\ $99.17$ \\ $-3.6\pm28.0$ \end{tabular} 
				& \begin{tabular}{c}$99.40$ \\ $99.10$ \\ $99.25 $ \\ $-4.1\pm35.3$ \end{tabular} \\
				
				\hline
				\begin{tabular}{l} This work \\ (each lead is \\ used separately) \end{tabular}
			      & \begin{tabular}{c}{\it Se} (\%) \\ {\it PPV} (\%) \\ {\it F}1 (\%) \\ $m\pm\sigma(ms)$ \end{tabular} 
				    &	\begin{tabular}{c}$97.38$ \\ $95.53$ \\ $96.47$ \\ $0.9\pm14.1$ \end{tabular} 
						& \begin{tabular}{c}$97.36$ \\ $95.52$ \\ $96.43$ \\ $-3.5\pm15.7$ \end{tabular} 
						& \begin{tabular}{c}$99.96$ \\ $99.84$ \\ $99.90$ \\ $2.1\pm9.8$ \end{tabular} 
						& \begin{tabular}{c}$99.96$ \\ $99.84$ \\ $99.90$ \\ $1.6\pm9.8$ \end{tabular} 
						& \begin{tabular}{c}$99.43$ \\ $98.88$ \\ $99.15$ \\ $1.3\pm 20.9$ \end{tabular} 
						& \begin{tabular}{c}$99.48$ \\ $98.94$ \\ $99.21$ \\ $-0.3\pm22.9$ \end{tabular} \\
				
				\hline
		\end{tabular}}
	\end{center}
	\caption{The comparison of ECG segmentation algorithms}
	\label{table:Compare}
\end{table}

Analyzing the results, we can draw the following conclusions:
\begin{itemize}
	\item the indicators {\it Se} and {\it PPV} for the proposed algorithm are the most or almost the highest for all types of ECG segments; 
	\item averaging the answer over all $12$ leads helps to detect the complexes better: it has improved both {\it Se} and {\it PPV}; 
	      however, the detecting the onsets and the offsets worsens, which is indicated by the growth of $\sigma$ in all indicators; 
	\item to detect the QRS-complexes, it is enough to use only lead II, since it gives the highest quality of their determination; 
	      such an approach will reduce the time of the algorithm $12$ times, without passing the other leads through the neural network;
	\item the best $\sigma$ values are given by the algorithm \cite{Kalyakulina2018};
	\item the results of the proposed approach for all indicators surpassed the other neural network approach \cite{Sereda2018}.
\end{itemize}

\subsection{Examples of the resulting segmentations} \label{sec:Examples}

\begin{figure}
  \includegraphics[width=\linewidth]{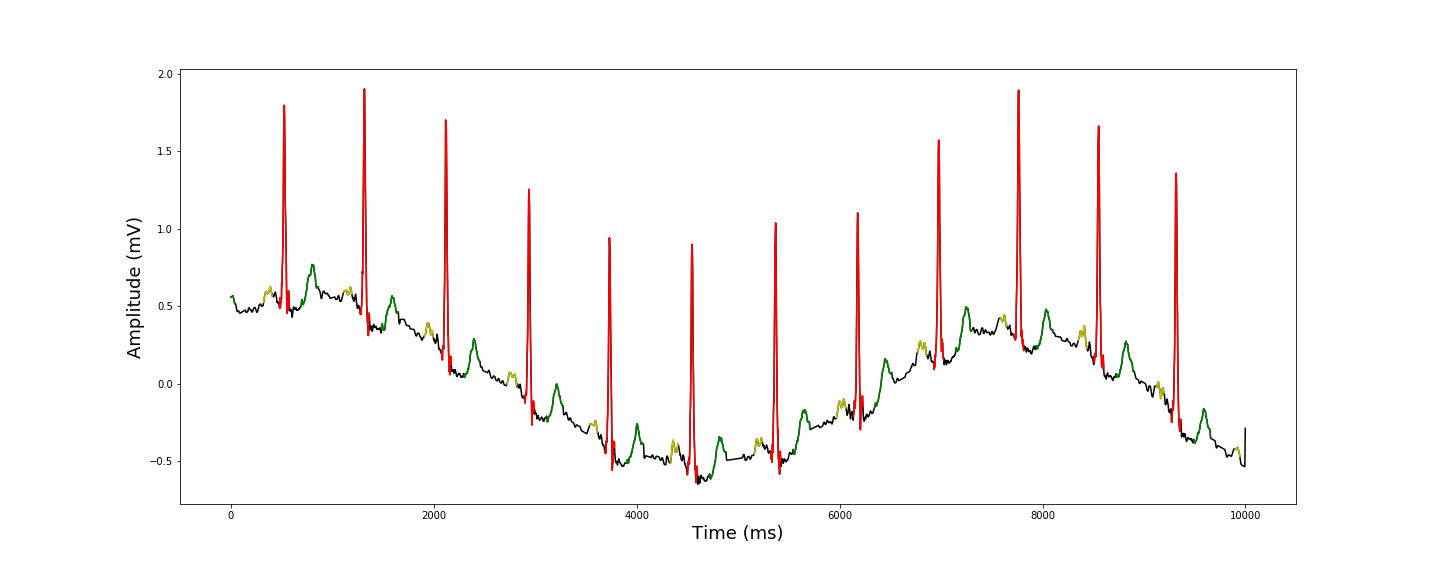}
  \caption{An example of low frequency noise ECG segmentation (breathing)}
  \label{fig:lNoise}
\end{figure}

\begin{figure}
  \includegraphics[width=\linewidth]{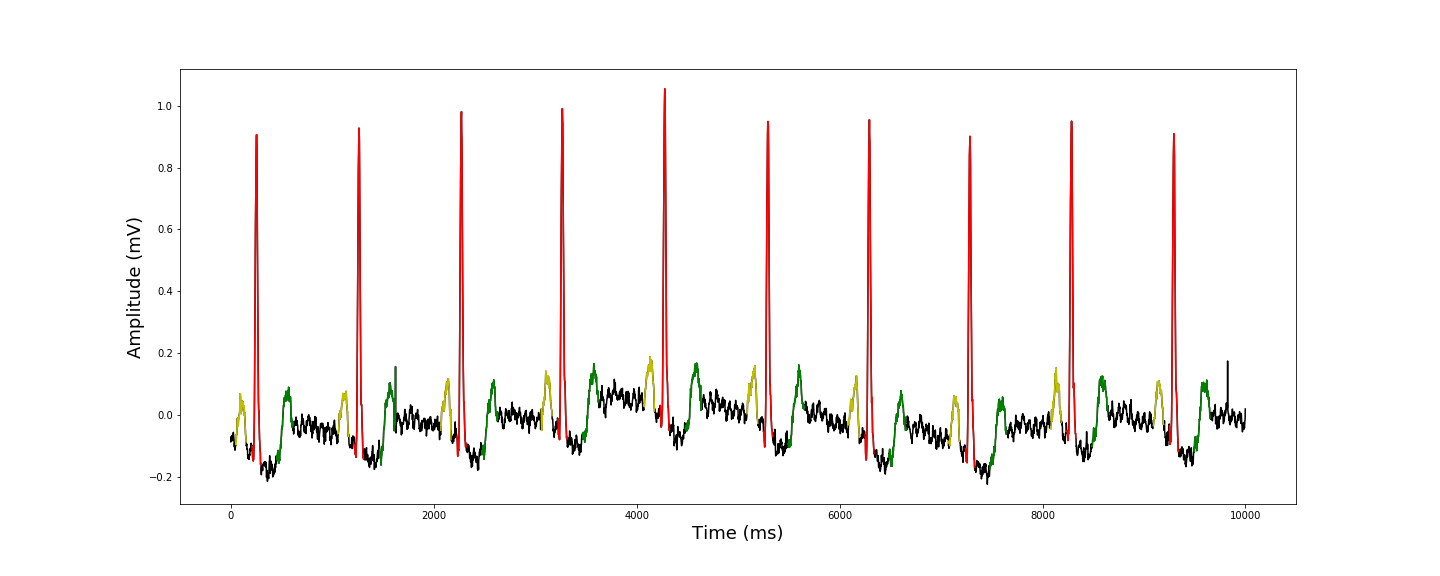}
  \caption{An example of high frequency noise ECG segmentation}
  \label{fig:hNoise}
\end{figure}

\begin{figure}
  \includegraphics[width=\linewidth]{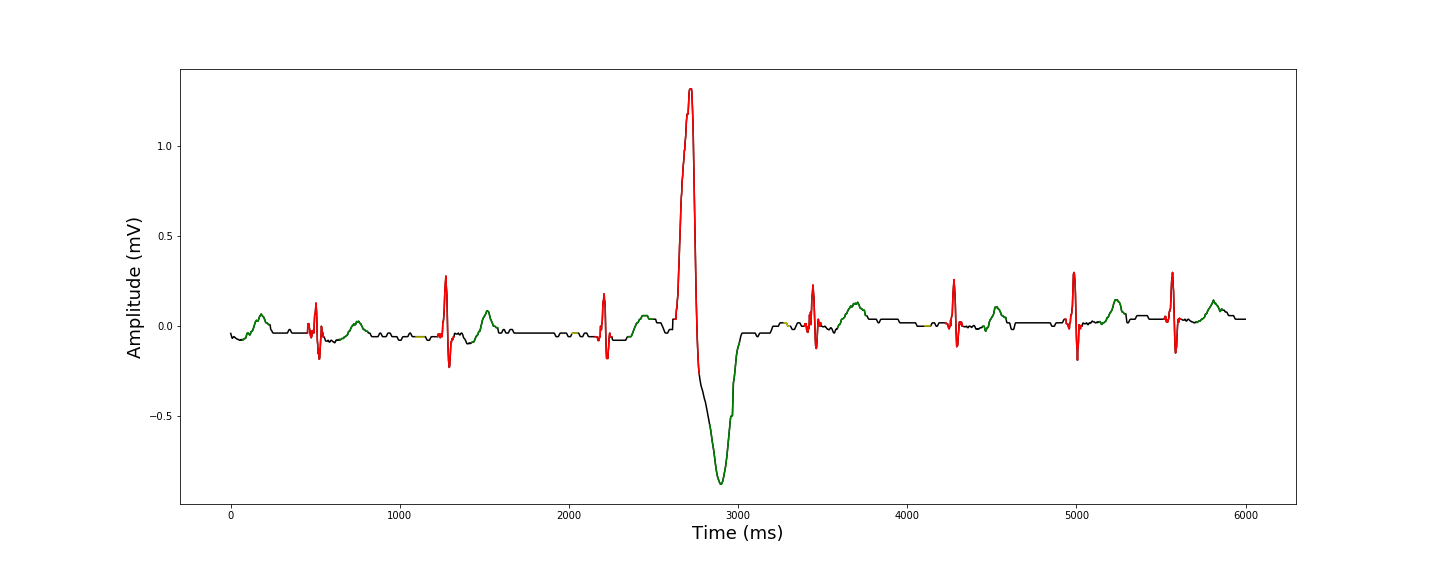}
  \caption{An example of ECG segmentation with pathology (ventricular extrasystole)}
  \label{fig:extra}
\end{figure}

\begin{figure}
  \includegraphics[width=\linewidth]{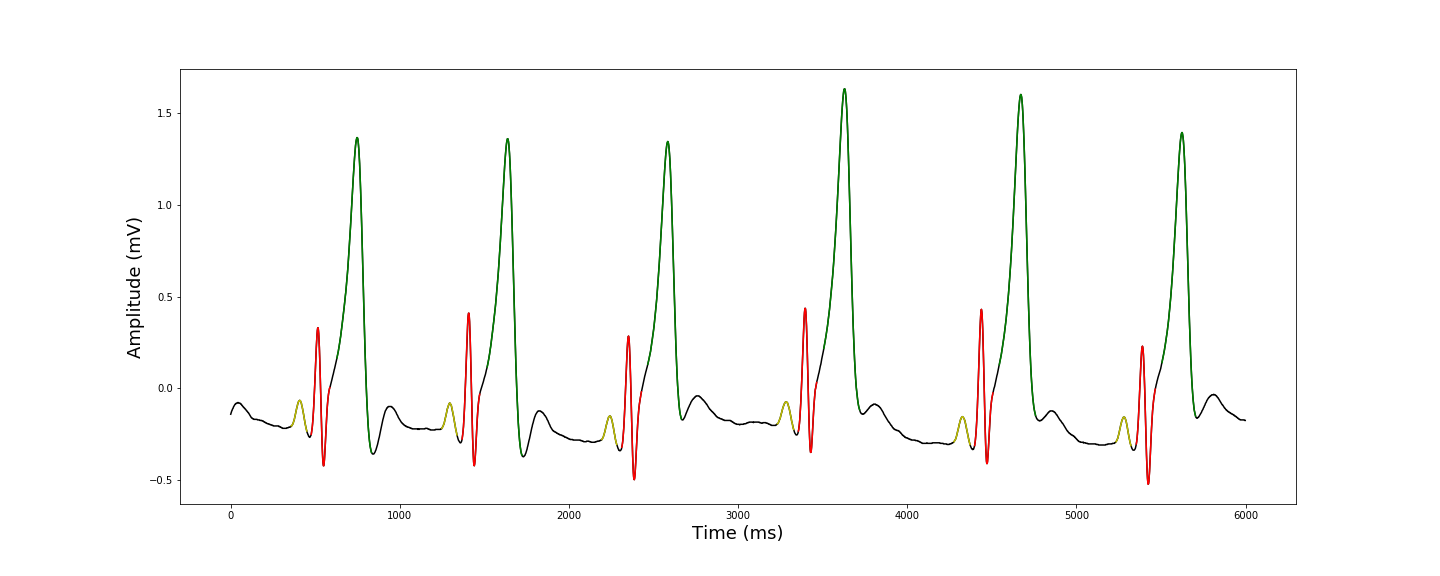}
  \caption{An example of segmentation of an ECG obtained from another type of ECG monitor. It is characterized by high T waves and a strong degree of smoothness.}
  \label{fig:anotherBase}
\end{figure}

\begin{figure}
  \includegraphics[width=\linewidth]{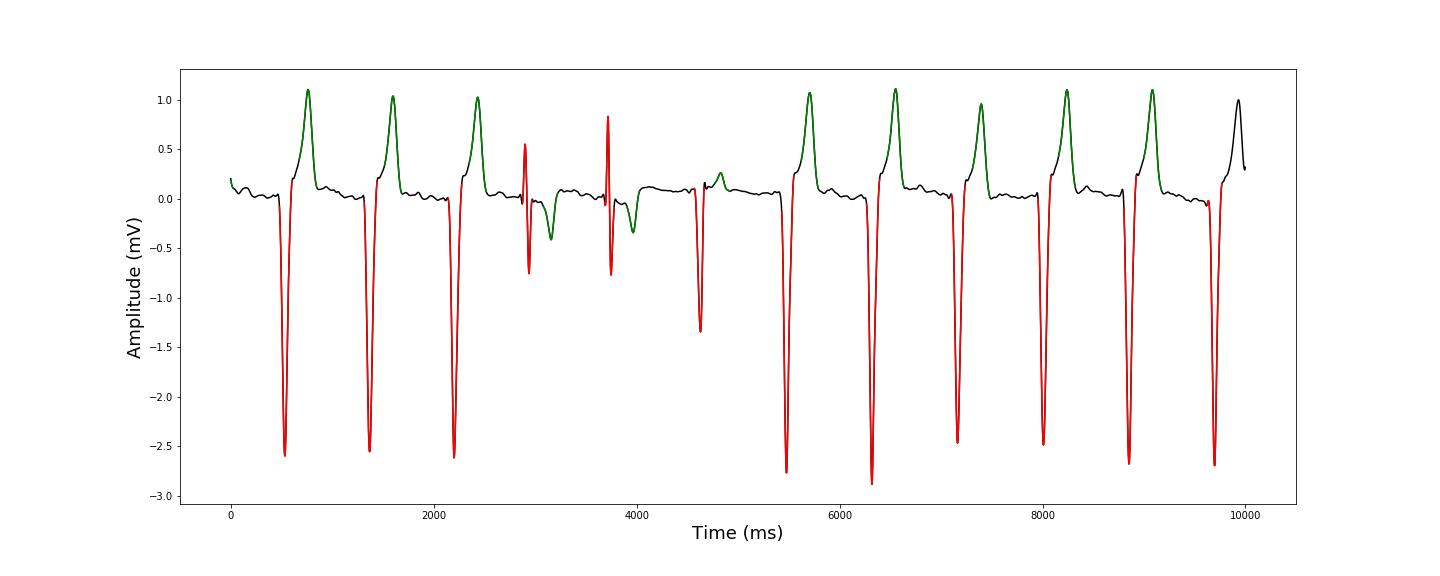}
  \caption{An example of segmentation of an ECG with the frequency of 50 Hz, reduced using a cubic spline to the frequency of 500 Hz}
  \label{fig:interpolate}
\end{figure}

Examples of segmentations obtained by the proposed algorithm are shown in Fig. \ref{fig:lNoise}--\ref{fig:interpolate}.

The experiments show that the proposed algorithm confidently copes with noise of different frequencies. 
An example with low frequency noise (breathing) is shown in Fig. \ref{fig:lNoise}.
An example with high frequency noise is presented in Fig. \ref{fig:hNoise}.

An example of the segmentation of an ECG with a pathology (ventricular extrasystole) is shown in Fig. \ref{fig:extra}.
An example of segmentation of an ECG obtained from another type of ECG monitor is shown in Fig. \ref{fig:anotherBase}. It is characterized by high T waves and a strong degree of smoothness.
Figure \ref{fig:interpolate} presents an example of segmentation of an ECG with the frequency of 50 Hz, reduced using a cubic spline to the frequency of 500 Hz.

\section{Conclusion and future work}

The paper describes an algorithm based on the use of a UNet-like neural network, which is capable to quickly and efficiently construct the ECG segmentation.
Our method uses a small number of parameters and it has a good generalization.
In particular, it is adaptive to different sampling rates and it is generalized to various types of ECG monitors.
The proposed approach is superior to other state-of-the-art segmentation methods in terms of quality. 
$F$1-measures for detection of onsets and offsets
of P and T waves and for QRS-complexes are at least $97.8$\%, $99.5$\%, and $99.9$\%, respectively.

In the future, this can be used with diagnostic purposes. Using segmentation, one can compute useful signal characteristics or use the neural network output directly as a new network input for automated diagnostics with the hope of improving the quality of classification.

In addition, one can try to improve the algorithm itself. In particular, the loss function used in the proposed neural network probably does not quite reflect the quality of segmentation. For example, it does not take into account some features of the ECG (e.\,g. two adjacent QRS complexes cannot be too close to each other or too far from each other). 

\paragraph{Acknowledgement.}
The authors are grateful to the referee for valuable suggestions and comments. 
The work is supported by the Ministry of Education and Science of Russian Federation (project 14.Y26.31.0022).

\end{document}